\documentclass[twocolumn,showpacs,preprintnumbers,amsmath,amssymb]{revtex4}

\usepackage{graphicx}% Include figure files
\usepackage{dcolumn}% Align table columns on decimal point
\usepackage{bm}% bold math
\usepackage{mathrsfs}
\usepackage{color}

\begin{document}

\title{Novel and simple description for a smooth transition from $\alpha$-cluster wave functions to $jj$-coupling shell model wave functions}% 

\author{Tadahiro Suhara$^1$, Naoyuki Itagaki$^2$, J\'{o}zsef Cseh$^3$, and Marek P{\l}oszajczak$^4$}

 \affiliation{
$^1$Institute of Physics, University of Tsukuba, Tsukuba 305-8571, Japan\\
$^2$Yukawa Institute for Theoretical Physics, Kyoto University,
Kitashirakawa Oiwake-Cho, Kyoto 606-8502, Japan\\
$^3$Institute of Nuclear Research of the Hungarian Academy of Sciences, Debrecen, Pf. 51, Hungary-4001\\
$^4$Grand Acc\'{e}l\'{e}rateur National d'Ions Lourds (GANIL), CEA/DSM -- CNRS/IN2P3, 
BP 5027, F-14076 Caen Cedex 05, France}

\date{\today}

\begin{abstract}
We propose an improved version of Antisymmetrized Quasi-Cluster Model (AQCM) 
to describe a smooth transition from the $\alpha$-cluster wave function to the $jj$-coupling shell model
wave function and apply it to the ground state of $^{12}$C. 
The cluster-shell transition in $^{12}$C is characterized in AQCM by only two parameters: 
$R$ representing the distance between $\alpha$ clusters and the center of mass, 
and $\Lambda$ describing the break of $\alpha$ clusters. 
The optimal AQCM wave function for the ground state of $^{12}$C is 
an intermediate state between the three-$\alpha$ cluster state and the shell model state with 
the $p_{3/2}$ subshell closure configuration. 
The result is consistent with that of the Antisymmetrized Molecular Dynamics (AMD),
and the optimal AQCM wave function quantitatively agrees with the AMD one, 
although the number of degrees of freedom in AQCM is significantly fewer.
\end{abstract}

\maketitle

\section{Introduction}

Shell structure is a fundamental property of atomic nuclei. The stability of nuclei and presence of magic gaps is explained by non-uniformities of the single-particle level distribution \cite{balian}. The strong spin-orbit interaction is essential to explain the observed magic numbers \cite{Mayer,Jensen}. 
The nuclear shell model in which the one-body behavior is supplemented by configuration-mixing effects of residual two-body interaction, describes nucleus as a closed quantum system where nucleons occupying bound orbits are isolated from the environment of scattering states and decay channels. In its modern version, shell model calculates nuclear properties in \textit{ab initio} manner \cite{peter}, using realistic interactions which reproduce the nucleon-nucleon scattering data \cite{rev1,rev2}.

The validity of such a closed quantum system framework depends on the dissociation energy. The configuration-mixing effects in weakly bound or unbound nuclear states cannot be treated as a small perturbation atop the mean field, and involve effects of the coupling to decay channels \cite{doba06}. In particular, the appearance of cluster states in the vicinity of their respective cluster decay thresholds is a consequence of an openness of the nuclear many-body system \cite{kyoto,forsch}.  Consequently, the standard shell model approach simply cannot be successful in predicting cluster states at low excitation energies around the cluster-decay thresholds. In the most advanced closed quantum system approaches to cluster decay, the shell model wave functions must be supplemented with a cluster component to achieve a quantitative agreement with the data \cite{rf:cmsm1}. 

The failure of the closed quantum system approaches to describe cluster states is the central problem in nuclear theory. The energetic order of particle emission thresholds, and their nature, depends on precise properties of the nuclear Hamiltonian. 
On the other hand, the phenomenological rule that cluster correlations are seen only in the vicinity of the respective cluster emission threshold cannot be a consequence of specific properties of nuclear forces.  A generic explanation of this rule in terms of  the collective external coupling of shell model states via the decay channel(s) has been put forward in Refs. \cite{kyoto,forsch}. In this context, $\alpha$-cluster states are of particular interest because of strong binding of an $\alpha$ particle and a weak $\alpha$-$\alpha$ interaction which does not allow to bind an $\alpha$-$\alpha$ system. The systematics of $\alpha$-cluster states in light nuclei was a basis for Ikeda conjecture that $\alpha$-cluster states appear close to their cluster decay thresholds \cite{Ike68}. 

In spite of recent remarkable advances in the open quantum system formulation of the shell model \cite{rf:8,rf:2}, a coherent picture of shell structure and clustering in the framework is not yet within our grasp. The closed quantum system description of low-energy excitations in light nuclei, where cluster- and shell-model-structures are intertwined, requires a generalization of the shell model wave function.

On the other hand, cluster models have been developed and successfully applied to describe cluster states. 
In traditional cluster models, each $\alpha$ cluster is expressed as $(s_{1/2})^{4}$ shell model configuration.
In this case, the spin-orbit interaction, which plays an essential role in the nuclear systems, 
cannot be taken into account.
This failure is coming from the special symmetry of the wave function.
Since four nucleons shares the same spatial wave function, 
each $\alpha$ cluster is a spin-zero system, and the spin-orbit interaction does not work. 
If we take the limit of zero distance among the clusters,
the cluster model wave function agrees with the shell model one in the case of closed shell nuclei such as $^{16}$O and $^{40}$Ca.
However, it cannot describe a subshell closure configuration, where the contribution of the spin-obit interaction becomes maximum.
Therefore, extension of the cluster model space, especially for the spin configurations, is needed 
for the general description of the nuclear structure.
In recent years, there have been many microscopic studies along this line, 
and competition of cluster and shell components in the 
ground state of light nuclei have investigated \cite{CS,SSS}.
For example, the ground state of $^{12}$C is an intermediate state 
between the three-$\alpha$ cluster state and 
the shell model state with the $p_{3/2}$ subshell closed configuration \cite{En'yo_12C_98,CS,Neff_12C_04,En'yo_12C_07}.
However it is still difficult to evaluate quantitatively 
to what extent cluster structures develop or shell model structures admix. 

A step forward in this direction is proposed in the Antisymmetrized Quasi-Cluster Model (AQCM) \cite{Simple,Masui,Yoshida,Ne-Mg} which attempts to include these two distinct structures of different physical origins in a single many-body approach.
In the AQCM, the transition from the cluster- to shell-model-structure can be described by only two parameters: $R$ representing the distance between $\alpha$ clusters and the center of mass, and $\Lambda$ which characterizes the quasi-cluster(s) and quantifies the role of the spin-orbit interaction in breaking the $\alpha$ cluster(s).
This is very transparent and simple approach to quantitatively discuss the mixing of cluster and shell components.
However, the previous AQCM, which were applied to the cluster-shell competition 
in C, Ne and Mg isotopes \cite{Simple,Masui,Yoshida,Ne-Mg,YKIS}, has a problem;
description for the subshell closure configurations was not exact. 
For example, in the studies of C isotopes \cite{Simple,Masui}, only one of the three $\alpha$ clusters was changed 
into a quasi-$\alpha$ cluster which corresponds to the $p_{3/2,\pm 3/2}$ shell model orbits.
The $p_{3/2,\pm 1/2}$ shell model orbits were not included in the model space because the remaining $\alpha$ clusters were unchanged.
The purpose of the present work is to improve the AQCM description 
by removing the restriction for the spin orientations of individual nucleons. 
In the new formulation, \textit{all} $\alpha$ clusters can be changed into quasi-$\alpha$ clusters, 
and $jj$-coupling shell model states including the subshell closure configuration can be described.

The paper is organized as follows. 
The formulation and details of the improved AQCM parameterization are given in Sect.~\ref{model}. 
In Appendix~\ref{subshell-closure}, we prove that this AQCM can describe the $p_{3/2}$ subshell closure configuration. 
In Sect.~\ref{results}, the AQCM results for the ground state of $^{12}$C are discussed and compared with 
the results of the Antisymmetrized Molecular Dynamics (AMD). 
We also briefly discuss the structure of the $0^{+}_{2}$ state. 
Finally, in Sect.~\ref{summary} we summarize the results and give the main conclusions.

\section{The model}\label{model}

In this section, we discuss the improved parameterization of the AQCM wave function and the many-body Hamiltonian used in this work.

\subsection{Single particle wave function (Brink model)}

In conventional $\alpha$ cluster models, the single particle wave function is described as
a Gaussian packet \cite{Brink}:
\begin{equation}
	\phi_{i} = \left( \frac{2\nu}{\pi} \right)^{\frac{3}{4}}
	\exp \left[- \nu \left(\bm{r}_{i} - \bm{R}_{\gamma} \right)^{2} \right] \eta_{i},
\end{equation}
where $\eta_{i}$ represents the spin-isospin part of the wave function, 
and $\bm{R}_{\gamma}$ is a real parameter representing the center of a Gaussian for the $\gamma$-th $\alpha$ cluster. 
In this Brink-Bloch wave function, four nucleons in the $\gamma$-th $\alpha$ cluster share the common $\bm{R}_{\gamma}$ value. 
Hence, the contribution of the spin-orbit interaction vanishes. 

\subsection{Single particle wave function in the AQCM}

In the AQCM, a nucleus consists of quasi-$\alpha$ clusters. 
For nucleons in the quasi-$\alpha$ cluster, the single particle wave function is described by a Gaussian packet, 
in the same way as in the Brink-Bloch wave function.
However, the center of this packet $\bm{\zeta}_{i}$ is a complex parameter
\begin{align}
	\psi_{i} &= \left( \frac{2\nu}{\pi} \right)^{\frac{3}{4}}
		\exp \left[- \nu \left(\bm{r}_{i} - \bm{\zeta}_{i} \right)^{2} \right] \chi_{i} \tau_{i}, \label{AQCM_sp} \\
	\bm{\zeta}_{i} &= \bm{R}_{\gamma} + i \Lambda \bm{e}^{\text{spin}}_{i} \times \bm{R}_{\gamma}. \label{center}
\end{align}
$\chi_{i}$ and $\tau_{i}$ in Eq.~\eqref{AQCM_sp} represent the spin and isospin part of the $i$-th single particle wave function, respectively. 
For the width parameter, we use the value $\nu = 0.235$ fm$^{-2}$.
The spin orientation is given by the parameter $\bm{\xi}_{i}$, while the isospin part is fixed to be 'up' (proton) or 'down' (neutron),
\begin{align}
	\chi_{i} &= \xi_{i\uparrow} |\uparrow \ \rangle + \xi_{i\downarrow} |\downarrow \ \rangle,\\
	\tau_{i} &= |p \rangle \ \text{or} \ |n \rangle.
\end{align}
In Eq.~\eqref{center}, $\bm{e}^{\text{spin}}_{i}$ is a unit vector for the intrinsic-spin orientation, 
and $\Lambda$ is a real control parameter describing the dissolution of the (quasi)-$\alpha$ cluster. 
As one can see immediately, the $\Lambda = 0$ AQCM wave function, which has no imaginary part, 
is the same as the conventional Brink-Bloch wave function.
We explain later that the AQCM wave function corresponds to the $jj$-coupling shell model wave function
when $\Lambda = 1$ and $\bm{R}_{\gamma} \rightarrow 0$,
and the improved AQCM approach can describe the subshell closure configuration.

The spin-orbit interaction is intuitively interpreted as 
\begin{equation}
	\bm{l} \cdot \bm{s} = \left( \bm{r} \times \bm{p} \right) \cdot \bm{s} 
	= \left( \bm{s} \times \bm{r} \right) \cdot \bm{p}
\end{equation}
where $\bm{r}$, $\bm{p}$, and $\bm{s}$ are the position, the momentum, and the spin of the nucleon, respectively.
If nucleons have the momentum components parallel (anti-parallel) to $\bm{s} \times \bm{r}$, 
the spin-orbit interaction acts attractively (repulsively).
An imaginary part of a Gaussian wave packet means the momentum of nucleon:
\begin{equation}
	\langle \psi_{i} | \hat{\bm{p}}_{i} | \psi_{i} \rangle = 2 \hbar \sqrt{\nu} \text{Im}[\bm{\zeta}_{i}].
\end{equation}
Therefore, the AQCM wave function has the momentum parallel or anti-parallel to $\bm{s} \times \bm{r}$ 
by introducing this particular form ($\Lambda \bm{e}^{\text{spin}}_{i} \times \bm{R}_{\gamma}$) of the imaginary part 
in Eq.~\eqref{center}. 
$\Lambda > 0$ and $\Lambda < 0$ correspond to parallel and anti-parallel momenta, respectively.
In actual calculation, of course, the spin-orbit interaction is 
a two-body force, which is a function of $\bm{r}_{i}-\bm{r}_j$ and not of $\bm{r}$.

In this paper we focus on $^{12}$C, which consists of three $\alpha$ clusters,
and we assume that they are placed with an equilateral triangular configuration. 
For the first quasi-$\alpha$-cluster, the spin direction is defined along the $z$-axis
and the center of mass is set in the $x$ direction as $R \bm{e}_{x}$. 
Then we introduce imaginary parts of centers of Gaussians related to the momenta of nucleons as in Eq.~\eqref{center}, 
which means in the $y$ direction.
Hence, the centers of Gaussian wave packets are 
\begin{equation}
	\bm{\zeta}_{i} = R (\bm{e}_x + i \Lambda \bm{e}_y)
	\label{parameter_first_up}
\end{equation}
and
\begin{equation}
	\bm{\zeta}_{i} = R (\bm{e}_x - i \Lambda \bm{e}_y)
	\label{parameter_first_down}
\end{equation}
for the spin-up and spin-down nucleons, respectively.
Here, $\bm{e}_x$ and $\bm{e}_y$ are unit vectors on $x$ and $y$ axes, respectively. 

For the second and third quasi-$\alpha$-clusters, we construct their spin directions and
the centers of Gaussian wave packets by rotating the first quasi-$\alpha$ cluster. 
Since both the spatial and spin parts of the wave function are rotated simultaneously,
the relative angles among $\bm{r}$, $\bm{p}$, and $\bm{s}$ in the first quasi-$\alpha$-cluster are kept 
in the second and third quasi-$\alpha$ clusters. 
Thus, the spin-orbit interaction also acts in these quasi-$\alpha$ clusters as well as in the first one. 
We rotate both the spatial and spin parts of the first quasi-$\alpha$-cluster about the $y$-axis as
\begin{align}
	\psi_{i+4} &= \hat{R}(\alpha = 0, \beta = \theta_{1}, \gamma = 0) \psi_{i}, \\
	\psi_{i+8} &= \hat{R}(\alpha = 0, \beta = \theta_{2}, \gamma = 0) \psi_{i},
\end{align}
where $i=1-4$, $(\alpha, \beta, \gamma) = \Omega$ are the Euler angles, 
$\hat{R}(\Omega) = e^{- i \alpha \hat{J}_z} e^{- i \beta \hat{J}_y} e^{- i \gamma \hat{J}_z}$ is the rotation operator,
and $\theta_{1}$ and $\theta_{2}$ are rotational angles,
$\theta_{1} = 2 \pi / 3$ and $\theta_{2} = 4 \pi / 3$.
From the discussion in the next subsection,
it is understood that the AQCM wave function coincides with the $(s_{1/2})^4(p_{3/2})^8$ configuration 
under the proper $\theta_{1}$, $\theta_{2}$ values and the conditions of $\Lambda = 1$ and $R \rightarrow 0$.

The essential difference between the previous and present AQCM comes from the treatment of the spin orientation. 
In the early version of the AQCM \cite{Simple,Masui,Yoshida}, the intrinsic spin of each nucleon was quantized with 
respect to the $z$-axis and the spin direction was restricted to $z$ or $-z$ direction for all nucleons
\begin{equation}
	\bm{\xi} = (1, 0) \ \text{or} \ (0, 1)
\end{equation}
Therefore, it was impossible to change all $\alpha$ clusters to quasi-$\alpha$ clusters and describe the subshell closure configuration.

In the improved AQCM, we expand the model space and spin directions can be oriented in any direction.
This means $\xi_{i\uparrow}$ and $\xi_{i\downarrow}$ are complex parameters
depending on the rotational angles $\theta_{1}$ and $\theta_{2}$.
As a result, we can change all $\alpha$ clusters to quasi-$\alpha$ clusters 
by rotating both the spatial and spin parts of the wave function of the first quasi-$\alpha$-cluster,
and describe the $p_{3/2}$ subshell closure configuration.

\subsubsection{$p_{3/2}$ subshell closure configuration}

Let us now discuss the AQCM single particle wave function analytically to prove that it describes the $p_{3/2}$ subshell closure configuration at $R \rightarrow 0$ and $\Lambda = 1$.

First, we discuss the spin-up nucleons in the first quasi-$\alpha$ cluster, 
whose centers of Gaussian wave packets are given in \eqref{parameter_first_up}. 
The single particle wave function is: 
\begin{equation}
	\psi_i = \left( \frac{2\nu}{\pi} \right)^{\frac{3}{4}}
		\exp \left[ -\nu \bm{r}_{i}^{2} -\nu \bm{\zeta}_{i}^{2} + 2\nu \bm{r}_{i} \cdot \bm{\zeta}_{i} \right] 
		|\uparrow \ \rangle \tau_{i}. 
\end{equation}
Using Eq. \eqref{parameter_first_up}, the last factor can be expanded as
\begin{align}
	\exp \left[ 2\nu \bm{r}_{i} \cdot \bm{\zeta} \right] 
		&= \sum_{k=0}^{\infty} \frac{1}{k!} (2 \nu R (x_{i} + i \Lambda y_{i}))^{k}, \nonumber\\
		&= \sum_{k=0}^{\infty} \frac{1}{k!} (2 \nu R r_{i})^{k} 
			\left( \frac{x_{i} + i \Lambda y_{i}}{r_{i}} \right)^{k}.
\end{align}
For $\Lambda = 1$, using the relations
\begin{equation}
	\left( \frac{x_{i} + i y_{i}}{r_{i}} \right)^{k} = \frac{1}{s_{k}} Y_{kk}(\Omega_{i}),
\end{equation}
and
\begin{equation}
	r_{i}^{k} \exp \left[ - \nu \bm{r}_{i}^{2} \right] = \frac{1}{t_{k}} R_{0k}(r_{i}),
\end{equation}
the single particle wave function becomes 
\begin{equation}
	\psi_{i} = \left( \frac{2\nu}{\pi} \right)^{\frac{3}{4}} 
		\sum_{k=0}^{\infty} \frac{(2 \nu R)^{k}}{k! s_{k} t_{k}} 
		R_{0k}(r_{i}) Y_{kk}(\Omega_{i}) |\uparrow \ \rangle \tau_{i}.
\end{equation}
Here, $s_{k}$ and $t_{k}$ are the normalization factors of spherical harmonics $Y_{kk}(\Omega_{i})$
and the radial wave function $R_{0k}(r_{i})$, respectively. 
Since the spin is up, this wave function is described as a linear combination of
$|j, j_z=j \rangle = R_{0k}(r_{i}) Y_{kk}(\Omega_{i}) |\uparrow \ \rangle$ states where $j = k + 1/2$:
\begin{equation}
	\psi_{i} = \sum_{j=1/2}^{\infty} a_{j} R^{j-\frac{1}{2}} |j, j \rangle \tau_{i}.
	\label{expansion_up}
\end{equation}
$a_{j}$ in \eqref{expansion_up} is the coefficient for the $|j, j \rangle$ state with the separated factor of $R^{j-\frac{1}{2}}$. 
Analogously, the single particle wave function for the spin-down nucleons in the first quasi-$\alpha$ cluster can be written as
\begin{equation}
	\psi_{i} = \sum_{j=1/2}^{\infty} a_{-j} R^{j-\frac{1}{2}} |j, -j \rangle \tau_{i}.
	\label{expansion_down}
\end{equation}
This expression can be obtained by a time-reversal transformation of the spin-up wave function.

Next, we discuss nucleons in the second and third quasi-$\alpha$-clusters 
which are generated by multiplying the rotational operator for the first quasi-$\alpha$ cluster.
According to the definition of Wigner $D$ function,
\begin{equation}
	\langle j, k | \hat{R}(\alpha, \beta, \gamma) | j, m \rangle 
	= \exp[-i k \alpha] d^j_{km}(\beta) \exp[-i m \gamma] 
	\label{Dfunc-def}
\end{equation}
where $d^{j}_{km}(\beta)$ is a Wigner small $d$ function, the rotated $|j, j \rangle$ state can be expressed as
\begin{equation}
	\hat{R}(\alpha = 0, \beta = \theta, \gamma = 0) |j, j \rangle = 
		\sum_{m=-j}^{j} d^{j}_{mj}(\theta) |j, m \rangle.
	\label{rotation_jj}
\end{equation}
Thus, the single particle wave functions in the second and third quasi clusters, 
which are generated by rotating $|j, j \rangle$ in the first quasi-cluster about the $y$-axis, can be written as
\begin{align}
	&\hat{R}(\alpha = 0, \beta = \theta, \gamma = 0) \psi_{i} \notag \\
		&= \sum_{j=1/2}^{\infty} \sum_{m=-j}^{j} a_{j} R^{j-\frac{1}{2}} d^{j}_{mj}(\theta) |j, m \rangle \tau_{i}.
	\label{convert_single_particle_wave_function}
\end{align}
We perform this transformation also for the spin-down nucleons in the first quasi-$\alpha$ cluster,
which are expressed as linear combinations of $\{ |j, -j \rangle\}$.
The rotation mixes other $|j, m \rangle$ states than the $|j, \pm j \rangle$ states,
when $\theta$ satisfies $d^{j}_{\pm jj}(\theta) \ne 1$.

Rotating all single particle wave functions following Eq. \eqref{convert_single_particle_wave_function}
and substituting them into Eq. \eqref{total_system},
we can discuss the nature of the Slater determinant consisting of quasi-clusters.
Owing to the anti-symmetrization, the lowest order of the wave function in $R$ 
is $R^{8}$ ($R^4$ for both proton and neutron parts), and the coefficient for $R^{8}$ exactly corresponds to the 
$(s_{1/2})^4(p_{3/2})^8$ configuration. 
Therefore, for $\Lambda = 1$, $R \rightarrow 0$, and proper $\theta_{1}$ and $\theta_{2}$ values, 
the total wave function coincides with the $p_{3/2}$ subshell closure configuration 
(see Appendix \ref{subshell-closure}).
This limit is different from the $R \rightarrow 0$ limit of the three-$\alpha$ cluster model,
which is known as the Elliott SU(3) limit, $(s)^{4}(p_{x})^{4}(p_{y})^{4}$.
%In the following, we show numerically that the AQCM wave function can describe the $p_{3/2}$ subshell closure configuration.

\subsection{AQCM wave function of the total system}

The AQCM wave function of the total system consisting of three quasi-$\alpha$-clusters 
is projected onto parity and angular momentum eigenstates using the parity projection operator $\hat{P}^{\pm}$ 
and the angular momentum projection operator $\hat{P}^{J}_{MK}$ as
\begin{equation}
\Phi = \sum_{K} C_{K} \hat{P}^{J}_{MK} \hat{P}^{\pm}\mathscr{A} \left[\psi_{1}, \psi_{2} , \cdots , \psi_{12} \right]
\label{total_system}
\end{equation}
The parity projection operator $\hat{P}^{\pm}$ is defined as 
\begin{equation}
	\hat{P}^{\pm} \equiv \frac{1 \pm \hat{P}}{2},
\end{equation}
where $\hat{P}$ is the parity operator.
The angular-momentum projection operator $\hat{P}^{J}_{MK}$ is defined as
\begin{equation}
	\hat{P}^{J}_{MK} \equiv \frac{2 J + 1}{8 \pi^{2}} \int d \Omega D^{J*}_{MK}(\Omega) \hat{R}(\Omega),
\end{equation}
where $D^{J}_{MK}(\Omega)$ is the Wigner $D$ function. 
In the total wave function, Eq. \eqref{total_system}, particles whose numbers are from $4N-3$ to $4N$ 
form the $N$-th quasi-$\alpha$ cluster.

\subsection{Hamiltonian \label{interact}}

The Hamiltonian operator $\hat{H}$ has the following form:
\begin{equation}
	\hat{H} = \sum_{i=1}^{A} \hat{t}_i - \hat{T}_{\text{c.m.}} + \sum_{i>j}^{A} \hat{v}_{ij},
\end{equation}
where a two-body interaction $\hat{v}_{ij}$ includes the central part, the spin-orbit part and the Coulomb part. 
For the central part, we use the Volkov No.2 effective $N-N$ potential \cite{Vol}:
\begin{equation}
	V^{\text{c}}(\hat{r}_{ij}) = \sum_{k=1}^{2} V_{k}^{\text{c}} \exp(-\hat{r}_{ij}^{2} / c_{k}^2) (W - M \hat{P}^\sigma \hat{P}^\tau),
\end{equation}
where $M = 0.60$, $W = 1 - M = 0.40$, $V_{1}^{\text{c}} = -60.65$ MeV, $V_{2}^{\text{c}} = 61.14$ MeV, 
$c_{1} = 1.8$ fm, and $c_{2} = 1.01$ fm. 
For the spin-orbit potential, we adopted the spin-orbit part of the G3RS potential \cite{G3RS}:
\begin{equation}
	V^{ls}(\hat{r}_{ij}) = \sum_{k=1}^{2} V_{k}^{ls} \exp(-\hat{r}_{ij}^{2} / d_{k}^2) 
		\hat{P}(^3\text{O}) \hat{\bm{L}} \cdot \hat{\bm{S}},
\end{equation}
where $V_{1}^{ls} = -1600$ MeV, $V_{2}^{ls} = 1600$ MeV, $d_{1} = 0.600$ fm, $d_{2} = 0.477$ fm, 
and $\hat{P}(^3\text{O})$ is a projection operator onto a triplet-odd state. The operator $\hat{\bm{L}}$ stands for the relative angular momentum and $\hat{\bm{S}}$ is the spin operator.

The parameter set: $M = 0.60$, $V_{1}^{ls} = -2000$ MeV, and $V_{2}^{ls} = 2000$ MeV 
is known to give a reasonable description of $\alpha+n$ and $\alpha+\alpha$ scattering phase shifts \cite{Okabe79}. In the present study, we adopt, however, the slightly weaker strengths of the spin-orbit force 
to fit the $0^{+}_{1}$ energy of $^{12}$C \cite{Suhara_AMD_10}.

\section{Results and discussion}\label{results}

In this section, we show results of the AQCM for $^{12}$C and compare them with the AMD results. In particular, we extract the single particle orbits with the Antisymmetrized Quasi-Cluster + Hartree-Fock (AQC+HF) method to show that the improved AQCM can describe the $p_{3/2}$ subshell closure configuration. We also briefly discuss the structure of the $0^{+}_{2}$ state. 

\subsection{Energy surfaces}

\begin{figure}[t]
	\centering
	\includegraphics[width=8.6cm]{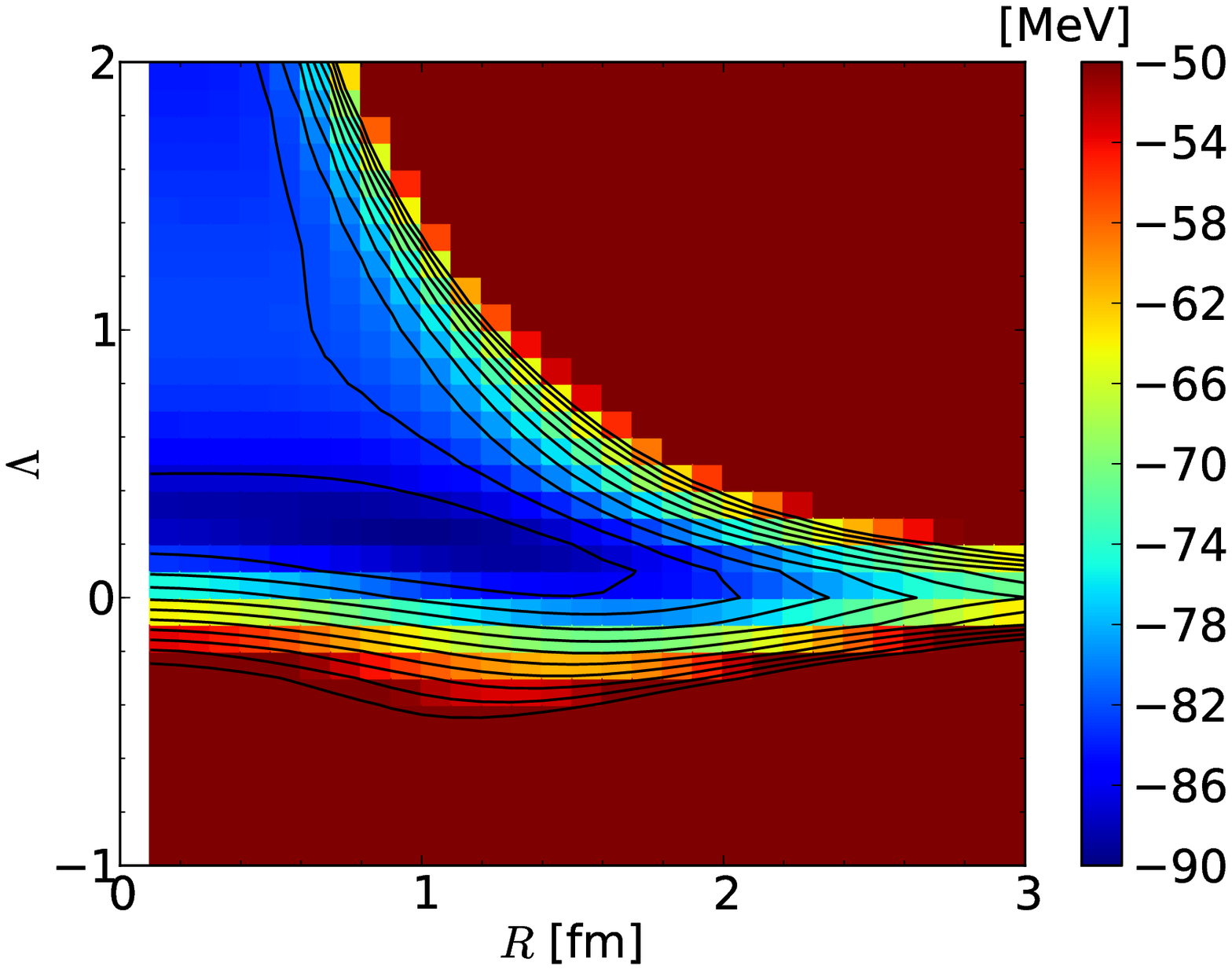} \\
	\includegraphics[width=8.6cm]{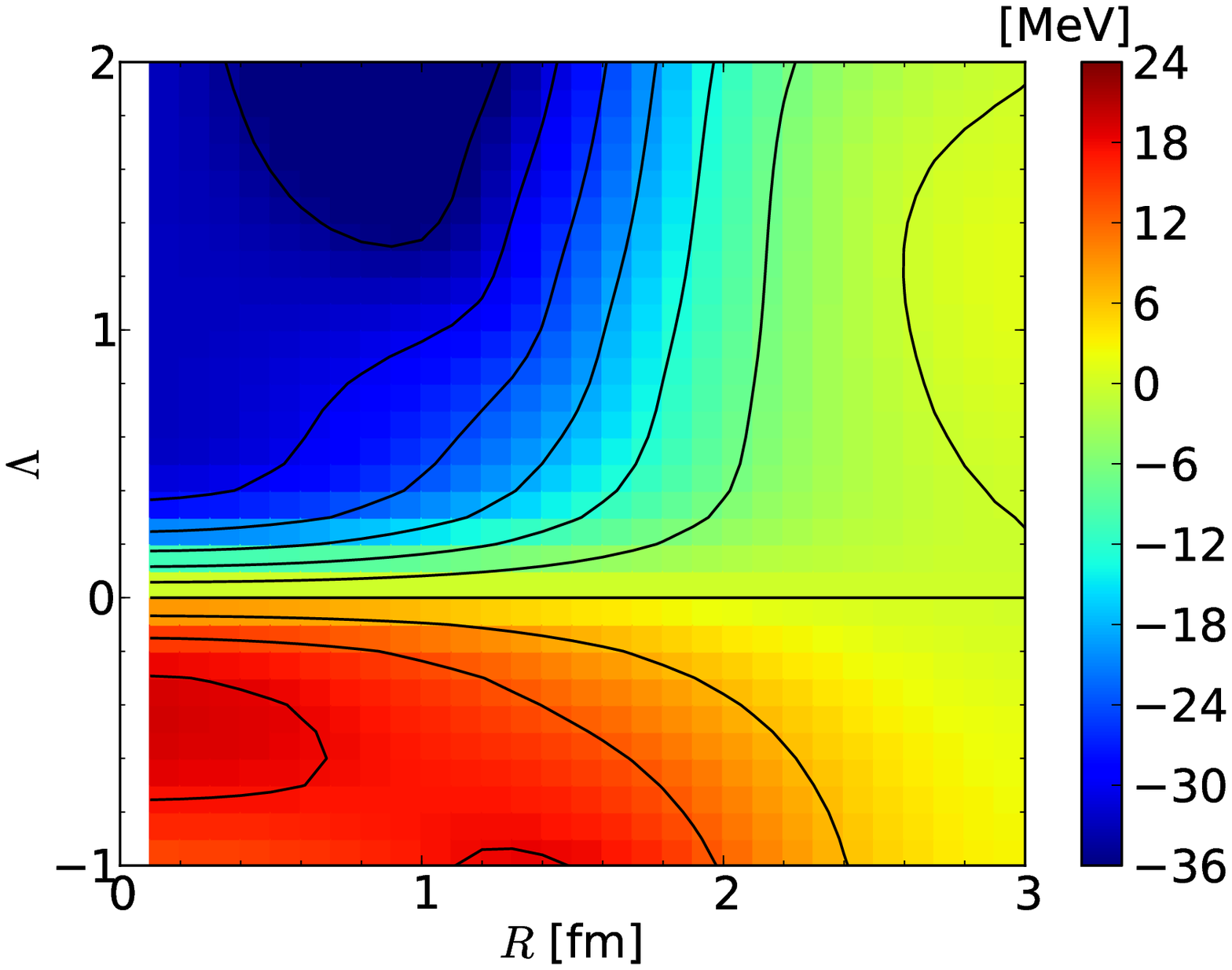} 
	\caption{(Color online) The energy surfaces for 
	the $0^{+}$ state of $^{12}$C, the total energy (top) and the spin-orbit energy (bottom), respectively.}
	\label{12C_surfaces}
\end{figure}

The AQCM (three quasi clusters) energy surfaces of $^{12}$C as functions of $R$ and $\Lambda$ are shown in Fig. \ref{12C_surfaces}. 
The top and bottom panels show the energy and spin-orbit energy surfaces
for the $0^{+}$ state of $^{12}$C, respectively. 
The minimum of the energy surface appears at around $(R, \Lambda) = (0.9, 0.2)$ and the corresponding energy is $- 89.6$ MeV.
This small distance between clusters, $R = 0.9$ fm, and a small but finite imaginary part of the wave function, $\Lambda = 0.2$,
indicate that the ground state of the $^{12}$C is an intermediate state between the three-$\alpha$ state ($\Lambda = 0$) 
and the $p_{3/2}$ subshell closure state ($R \rightarrow 0$ and $\Lambda = 1$). 
This hybrid character of the ground state wave function can be confirmed by calculating the squared overlap of the wave functions:
\begin{align}
	| \langle \Phi ((R, \Lambda) = (0.9, 0.0)) | \Phi_{\text{min}} \rangle |^{2} &= 63.6 \% , \\
	| \langle \Phi ((R, \Lambda) = (0.01, 1.0)) | \Phi_{\text{min}} \rangle |^{2} &= 47.3 \% ,
\end{align}
where $\Phi_{\text{min}}$ is the wave function at the minimum of the energy surface. One can see that
the squared overlaps between $\Phi_{\text{min}}$ and the three-$\alpha$ state ($(R, \Lambda) = (0.9, 0.0)$),
and between $\Phi_{\text{min}}$ and the $p_{3/2}$ subshell closure state ($(R, \Lambda) = (0.01, 1.0)$)
have both significant values around 0.5.

From the spin-orbit energy surface, one finds that the spin-orbit interaction acts attractively and repulsively for $\Lambda > 0$ 
and $\Lambda < 0$, respectively (for $\Lambda = 0$, the spin-orbit interaction does not act).
Around the minimum of the total energy, the spin-orbit interaction contributes $\sim -16$ MeV to the total energy.
In the next subsection, we will show that the improved AQCM takes into account effects of the spin-orbit interaction very well, and the AQCM wave function is almost the same as the AMD wave function.

\subsection{Single particle orbits of the AQC+HF method}\label{HF}

\begin{figure}[t]
	\centering
	\includegraphics[width=8.6cm]{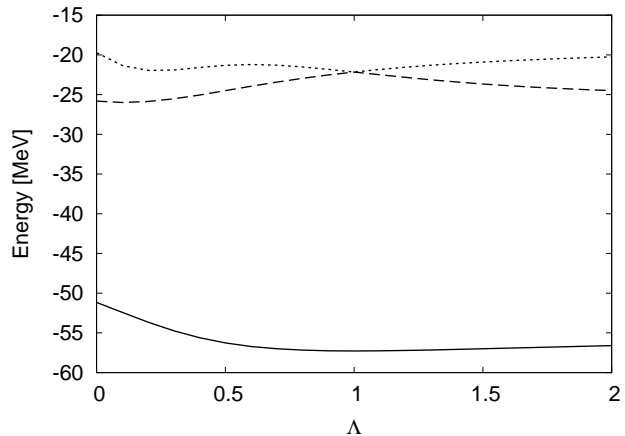}
	\caption{The AQC+HF effective single particle energies of the improved AQCM are plotted as a function of the parameter $\Lambda$ ($R = 0.01$ [fm]) in $^{12}$C. The Coulomb interaction is neglected in this calculation.}
	\label{12C_orbit_3}
\end{figure}

The AQCM single particle wave functions are given by Gaussian packets with complex parameters, 
and different single particle wave functions are mutually non-orthogonal. 
To extract orthogonal single particle orbits and corresponding effective single particle energies from AQCM, 
we apply the method imitating Hartree-Fock (HF) approach which was applied first 
in the context of the AMD \cite{Dote_AMDHF_97}.
We call it the AQC+HF method. 
In general, the corresponding AQC+HF single particle orbits are given by 
linear combinations of Gaussian single particle wave functions. 
One should stress however that the HF self-consistency in AQC+HF approach 
is satisfied only within the restricted functional space of AQCM single particle wave functions. 
Moreover, except for special cases, resulting single particle orbits are not 
eigenstates of angular momentum and parity operator.
In spite of these restrictions, the AQC+HF approach provides useful information 
about the dependence of mutually orthogonal single-particle orbits 
and effective single particle energies on parameters of the AQCM manifold. 
In the following, we analyze these effective single particle energies 
to show that the improved AQCM describes the $p_{3/2}$ subshell closure 
with the set of the parameter values $R \rightarrow 0$ and $\Lambda = 1$.

Fig.~\ref{12C_orbit_3} shows AQC+HF effective single particle energies of $^{12}$C 
for the improved AQCM as a function of $\Lambda$. 
The Coulomb interaction is neglected and the limit of $R \rightarrow 0$ is expressed by $R = 0.01$ fm. 
There are in this case only three AQC+HF single particle orbits expressed by solid, dashed, and dotted lines. 
Each of these orbits is occupied by four nucleons because of the spin-isospin degeneracy.
The lowest orbit (the solid line) corresponds to $s_{1/2}$ shell model wave function. 
The higher two orbits (dashed and dotted lines) are linear combinations of $p_{3/2}$ and $p_{1/2}$ components. 
At $\Lambda = 1$, these two single particle orbits become degenerate. 
In the vicinity of $\Lambda = 1$, the dominant component in higher orbits is $p_{3/2}$.
The $p_{1/2}$ component vanishes at $\Lambda = 1$. 
At this point, the higher single particle orbits are pure $p_{3/2}$ shell model wave functions. 
If we take into account the spin-isospin degeneracy of four, eight nucleons occupy these degenerate $p_{3/2}$ orbits.  
Hence, for $R \rightarrow 0$ and $\Lambda = 1$, the AQCM wave function describes 
the ground state configuration $(s_{1/2})^4 (p_{3/2})^8$ of $^{12}$C in the $jj$-coupling shell model. 
For other $\Lambda$ values, there are three independent AQC+HF single particle orbits. 
In this case, the AQCM wave function does not describe the $p_{3/2}$ subshell closure configuration.

\begin{figure}[t]
	\centering
	\includegraphics[width=8.6cm]{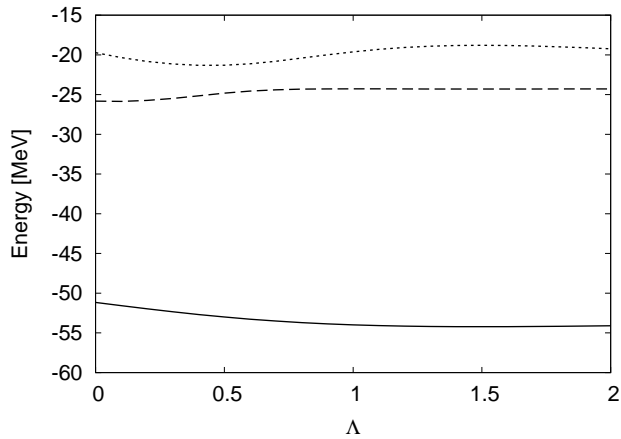}
	\caption{The same as in Fig. \ref{12C_orbit_3} but for the old version of the AQCM \cite{Simple,Masui,Yoshida}.}
	\label{12C_orbit_1}
\end{figure}

Fig. \ref{12C_orbit_1} shows the AQC+HF effective single particle of $^{12}$C 
for the old version of the AQCM \cite{Simple,Masui,Yoshida}, where only one $\alpha$ cluster is changed into a quasi-$\alpha$-cluster.
In this case, there exist always three independent AQC+HF single particle orbits for any value of $\Lambda$, 
\textit{i.e.} the old version of the AQCM cannot describe the $p_{3/2}$ subshell closure configuration. 

\begin{figure}[t]
	\centering
	\includegraphics[angle=0,width=8.6cm]{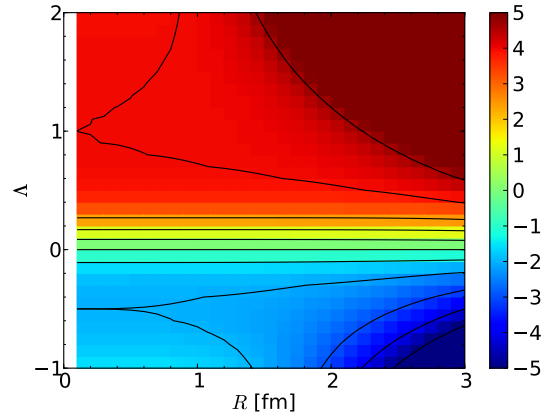}
	\caption{(Color online) The expectation value of one-body spin-orbit operator,
    $\sum_i \hat{\bm{L}}_{i} \cdot \hat{\bm{S}}_{i}$, in the improved AQCM.}
	\label{12C_obls}
\end{figure}

In Fig. \ref{12C_obls}, we show the expectation value of the one-body spin-orbit operator,
$\langle \Phi(R,\Lambda)|\sum_i \hat{\bm{L}}_{i} \cdot \hat{\bm{S}}_{i}|\Phi(R,\Lambda)\rangle$, as a function of $R$ and $\Lambda$. Here the sum over $i$ is for all nucleons. 
The calculations are performed using the improved AQCM. 
The expectation value becomes zero for the three-$\alpha$ model ($\Lambda = 0$), 
and its sign changes when crossing the $\Lambda = 0$ line. For the $p_{3/2}$ subshell closure configuration, the expectation value of the one-body spin-orbit operator should be 4, 
\textit{i.e.} 8 (number of nucleons in the $p_{3/2}$ orbit) $\times$ 1/2 
(the expectation value of $\hat{\bm{L}}_{i} \cdot \hat{\bm{S}}_{i}$ for the $p_{3/2}$ orbit). 
Indeed, we can see that this value is reached exactly in the limit of $\Lambda = 1$ and $R \to 0$. 
This proves that the AQCM can describe the $p_{3/2}$ subshell closure configuration.

The optimal AQCM state at $R = 0.9$ fm and $\Lambda$ = 0.2 which corresponds to the minimum of the energy surface, has the value $\langle \Phi(R,\Lambda)|\sum_i \hat{\bm{L}}_{i} \cdot \hat{\bm{S}}_{i}|\Phi(R,\Lambda)\rangle = 2.38$. Again, this shows that the main component in the ground state wave function is intermediate between the cluster ($\langle \Phi(R,\Lambda=0)|\sum_i \hat{\bm{L}}_{i} \cdot \hat{\bm{S}}_{i}|\Phi(R,\Lambda=0)\rangle = 0$) and $p_{3/2}$ subshell closure ($\langle \Phi(R\rightarrow 0,\Lambda=1)|\sum_i \hat{\bm{L}}_{i} \cdot \hat{\bm{S}}_{i}|\Phi(R\rightarrow 0,\Lambda=1)\rangle = 4$) limits.

\subsection{Comparison of AQCM and AMD results}\label{secamd}

In this subsection, we compare our results of the AQCM with AMD.
In Ref. \cite{Suhara_AMD_10}, we have calculated $^{12}$C using the $\beta$-$\gamma$ constrained AMD 
with the same interaction as used in the present work. 
In AMD, all nucleons are described as independent Gaussian wave packets, 
and the Gaussian center parameters are complex variational parameters. 
This variation called 'cooling process' is often performed before the angular momentum projection.
Here, we have introduced as constraints the quadrupole deformation parameters, $\beta$ and $\gamma$, 
and prepared many different states to solve the cooling equation. 
By projecting all these different configurations onto the eigenstates of angular momentum and parity, 
we can obtain the lowest energy configuration after the projection. 
Thus the $\beta$-$\gamma$ constrained AMD somehow overcomes the approximation of projection after variation 
and can be considered as an improved version of AMD.

We compare the optimal solution of our AQCM wave function $(R, \Lambda) = (0.9, 0.2)$ 
with this $\beta$-$\gamma$ constrained AMD at the minimum point of the $0^{+}$ energy surface on the $\beta$-$\gamma$ plane, 
$(\beta \cos \gamma , \beta \sin \gamma) = (0.35, 0.17)$, which is the dominant component of the $0^{+}_{1}$ state of $^{12}$C. 
The energy associated with these AQCM and AMD wave functions are $- 89.6$ MeV and $- 90.1$ MeV, respectively, 
\textit{i.e.} their difference is rather small, $0.5$ MeV.
In addition, the squared overlap between the AQCM and AMD wave functions is very large,
\begin{equation}
	| \langle \Phi_{\text{min}} | \Phi_{\text{AMD,{min}}} \rangle |^{2} = 98.6 \% ,
\end{equation}
where $\Phi_{\text{AMD,{min}}}$ is the $0^{+}$ projected AMD wave function 
at $(\beta \cos \gamma , \beta \sin \gamma) = (0.35, 0.17)$ in Ref. \cite{Suhara_AMD_10}.
This very large overlap indicates that the AQCM wave function gives 
an almost identical result of the AMD wave function for the ground state of $^{12}$C. 
Remaining small differences between AMD and AQCM wave functions are due to different symmetries in both models
as discussed later. 
The AQCM wave function has a threefold rotational symmetry about the $y$-axis, while the AMD wave function has no symmetry.

The number of degrees of freedom in AQCM and AMD wave functions is very different. 
The AQCM wave function has only two degrees of freedom ($R$ and $\Lambda$), 
whereas the AMD wave function in the studied case has $\sim 100$ degrees of freedom, 
such as real and imaginary parts of the center of Gaussian wave packet 
and the direction of the spin for each single particle wave function. 
This indicates that the AQCM describes effect of the spin-orbit interaction and the breaking of $\alpha$ clusters very efficiently.

In the framework of $\beta$-$\gamma$ constrained AMD, wave functions with the constraint of $\gamma = \pi/3$ 
have approximately threefold rotational symmetry. 
Therefore, one can compare AQCM and AMD wave functions having the same threefold symmetry 
by taking the minimum AMD wave function on the $\gamma = \pi/3$ line.
This minimum point is $(\beta, \gamma) = (0.29, \pi/3)$ and the corresponding energy is $- 89.6$ MeV, 
which is the same value of the optimal solution of the AQCM wave function. 
In addition, the squared overlap is very close to unity,
\begin{equation}
	| \langle \Phi_{\text{min}} | \Phi_{\text{AMD,sym}} \rangle |^{2} = 99.9 \% ,
\end{equation}
where $\Phi_{\text{AMD,sym}}$ is the $0^{+}$ projected AMD wave function at $(\beta, \gamma) = (0.29, \pi/3)$.
This almost 100 \% overlap of these two wave functions means that 
the AQCM result is consistent with the AMD result for the ground state of $^{12}$C
except for the effect of symmetry breaking.

Finally, we compare the $\beta$-$\gamma$ constrained AMD with the previous AQCM wave function \cite{Simple,Masui,Yoshida}, 
where only one of the $\alpha$ clusters is changed into the quasi-cluster.
In this version of the AQCM, the minimum point of the energy surface is at $(R, \Lambda) = (1.2, 0.4)$
and its energy is $-89.2$ MeV. 
Then, energy difference between the previous AQCM and AMD is $0.9$ MeV, 
which is not a serious problem compared with the improved AQCM, $0.5$ MeV.
However, the squared overlap of these two wave functions deviates very much from unity
\begin{equation}
	| \langle \Phi_{\text{pre,min}} | \Phi_{\text{AMD,{min}}} \rangle |^{2} = 90.8 \% ,
\end{equation}
where $\Phi_{\text{pre,min}}$ is the previous AQCM wave function at the minimum of the total energy.

\subsection{Comparison of cluster-shell competition in $0^{+}_{1}$ and $0^{+}_{2}$ states of $^{12}$C}

The first excited $0^{+}$ state in $^{12}$C is a well-known Hoyle resonance close to $\alpha$- and triple-$\alpha$-decay thresholds. The proximity of these two decay thresholds implies that the continuum coupling is particularly important for understanding of the structure of this state, similarly as the proximity of  one- and two-neutron decay channels determines the ground state properties of $^{11}$Li \cite{kyoto,forsch}. The present version of the AQCM describes many-body states in the closed quantum system framework, {\textit i.e.} the coupling of the wave function to the decay channels is absent and the asymptotic form of the wave function is incorrect. Keeping this shortcoming in mind, we employ the flexible AQCM to describe the evolution of cluster-shell competition from the ground state $0_{1}^{+}$ to the Hoyle state $0_{2}^{+}$.

Results of this subsection have been obtained using the Generator Coordinate Method (GCM) with a number of basis states. These states are given by AQCM wave functions for different $R$ and $\Lambda$ values, and various three-$\alpha$ cluster model wave functions. In these latter components, the AQCM assumption of  a threefold rotational symmetry about the $y$-axis is abandoned to simulate a gas-like near-threshold component of the wave function. 

\begin{figure}[t]
	\centering
	\includegraphics[angle=0,width=8.6cm]{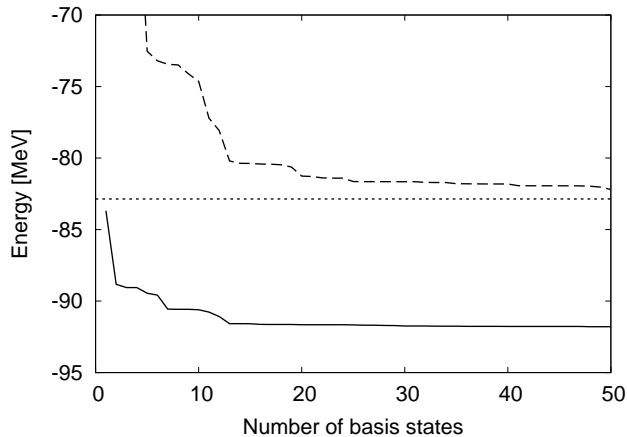}
		\caption{ The convergence of total energy for $0_1^{+}$ (the solid line) and $0_2^{+}$ 
		(the dashed line) states in $^{12}$C with respect          
		to the number of trial AQCM and three-$\alpha$ cluster basis states.
		The dotted line shows the three-$\alpha$ cluster threshold energy, -82.9 MeV.}
	\label{12C_conv}
\end{figure}

Results are shown in Fig.~\ref{12C_conv} for the ground state $0_{1}^{+}$ and the Hoyle state $0_{2}^{+}$. The dotted line in Fig.~\ref{12C_conv} shows the three-$\alpha$ cluster threshold energy which is $-82.9$ MeV in the present model. From the first to the tenth basis state, we superpose AQCM wave functions
where the values of parameters $R$ and $\Lambda$ are chosen randomly keeping a threefold rotational symmetry about the $y$-axis. The following forty basis states correspond to various three-$\alpha$ cluster wave functions which are obtained by assigning the Gaussian center positions of $\alpha$ clusters randomly. These basis states have no spatial symmetry.

Figure~\ref{12C_conv} shows the convergence of the total energy for $0_1^+$ (the solid line) and $0_2^+$ (the dashed line) states depending on the number of basis states. We see that the random three-$\alpha$ cluster component, which is inessential in the ground state of $^{12}$C, lowers the energy of the first excited state by about 7 MeV. Hence, the gas-like component cannot be neglected in the near-threshold $0_2^+$ state. Because of that, the energy of $0_2^+$ state converges rather slowly and one needs at least fifty basis states to find the stable result. The energy of the $0^{+}_{1}$ state converges at $-91.8$ MeV, which is close to the experimental value $-92.2$ MeV. The calculated $0^{+}_{2}$ state appears slightly above both $\alpha$- and triple-$\alpha$-decay thresholds, in agreement with the experimental data.

To discuss the structure of $0^{+}_{1}$ and $0^{+}_{2}$ states,
we calculate the expectation value of the one-body spin-orbit operator.
For the ground state, the calculated value, $\langle \Phi_{\rm GCM}(0_1^+)|\sum_i \hat{\bm{L}}_{i} \cdot \hat{\bm{S}}_{i}|\Phi_{\rm GCM}(0_1^+)\rangle = 1.77$, is in between 0 (the three-$\alpha$ cluster state limit)
and 4 (the $p_{3/2}$ subshell closure configuration limit).
This result confirms the earlier finding (see Sect. \ref{HF}) for a single optimal AQCM wave function that the $0^{+}_{1}$ state is 
intermediate between the three-$\alpha$ cluster state and the $p_{3/2}$ subshell closure state.
As compared to the value for an optimal state at $R = 0.9$ fm and $\Lambda$ = 0.2 (see Sect. \ref{HF}), $\langle \Phi(R,\Lambda)|\sum_i \hat{\bm{L}}_{i} \cdot \hat{\bm{S}}_{i}|\Phi(R,\Lambda)\rangle = 2.38$, the GCM value is slightly reduced due to the mixing of different basis states and breaking a threefold rotational symmetry about the $y$-axis.
On the other hand, the near-threshold $0^{+}_{2}$ state has a very small expectation value of the one-body spin-orbit operator,  $\langle \Phi_{\rm GCM}(0_2^+)|\sum_i \hat{\bm{L}}_{i} \cdot \hat{\bm{S}}_{i}|\Phi_{\rm GCM}(0_2^+)\rangle = 0.11$, {\textit i.e.} it is an almost pure cluster structure.

These results for $0_1^+$ and $0_2^+$ agree with the modern understanding of cluster structure formation in $^{12}$C \cite{kyoto, forsch}.
A transition between the $jj$-coupling shell model structure and the three-$\alpha$ cluster structure
occurs in the vicinity of the $\alpha$- and triple-$\alpha$-decay thresholds and can be seen in the near-threshold $0^{+}_{2}$ state. It is interesting that the present GCM calculation which neglects an explicit coupling to both $\alpha$-decay channels is capable to follow the transition from a predominantly shell model wave function in the ground state $0_1^+$ to an almost pure cluster wave function in $0^{+}_{2}$ excited state.

\section{Summary and perspectives}\label{summary}

Nuclear clustering escapes the description in terms of the standard shell model which simply fails to predict cluster states at observed low excitation energies around cluster-decay threshold. Generic explanation of this enigmatic phenomenon involves understanding of the role played by the coupling of shell model states via the decay channel(s). In this work, we have proposed a simple approach based on the improved AQCM parametrization to describe a transition from the $\alpha$-cluster wave function to the $jj$-coupling shell model wave function. As compared to the previous AQCM \cite{Simple,Masui,Yoshida,Ne-Mg} which allows 
the transformation of a single $\alpha$ cluster into a quasi-$\alpha$ cluster, 
in the improved AQCM we expand the model space and spin directions can be oriented in any direction.
In the new formulation, \textit{all} $\alpha$ clusters of a nucleus can be changed into quasi-$\alpha$ clusters.
Hence, a destructive interference of the spin-orbit coupling and clustering can be investigated in a single variational wave function. 

The relation between the shell and cluster models was established in the harmonic oscillator limit \cite{wika58}, 
\textit{i.e.} via the SU(3) symmetry \cite{babo58,elli58}. In the symmetry-adopted models, like SU(3) shell model \cite{elli58}, the symplectic shell model \cite{sympl}, or the semimicroscopic algebraic cluster model \cite{sacm}, one may characterize the relative importance of the shell and cluster components by the SU(3) content of the realistic wave function. The other way is to split the interaction into SU(3)-preserving and a SU(3)-breaking parts. Their relative weights can then be used as a control parameter to indicate how close the real situation is to the intersection point of the two models \cite{phase}. In the shell model, typical symmetry-preserving parts are the harmonic oscillator and quadrupole forces, while important symmetry-breaking parts are the spin-orbit and pairing interactions. In the cluster model, the symmetric parts are again the harmonic oscillator and quadrupole forces, and the symmetry-breaking interaction can be e.g. the dipole interaction. In the $0p$ shell, where the SU(3) model reduces to Wigner's supermultiplet theory \cite{wig37}, the joint conclusion of many works is that for the ground state of the $^{12}$C nucleus the $(0,4)$ SU(3) symmetry is a good approximation. The recent studies using the No-Core Shell Model \cite{nocore} confirms the dominance of the $(0,4)$ component in the wave function \cite{nocore12c}. 

The AQCM proposed in this work, provides a possibility for going from the molecule-like clusterization to the $jj$ coupled shell model limit directly, in a single step, contrary to the two-step procedure of the shell- and cluster-model calculations with different symmetry-breaking terms, as we discussed in Ref. \cite{Ne-Mg}. This model does not apply the SU(3) basis, thus a detailed comparison with the standard shell model is not easy. 
Based on analytical and numerical studies, we have shown that the AQCM wave function in the limit $R \rightarrow 0$ and $\Lambda = 1$ corresponds to the $(s_{1/2})^4 (p_{3/2})^8$ closed shell configuration of $^{12}$C. 
The proposed parametrization allows to determine an optimal wave function of $^{12}$C in a variational procedure 
for each chosen effective nucleon-nucleon interaction.  The optimal AQCM ground state of $^{12}$C is an intermediate state between the three-$\alpha$ cluster state and the shell model state with the $p_{3/2}$ subshell closure configuration. From a comparison with the AMD model, where all nucleons are treated independently, we found that the AQCM result is consistent with the AMD result (overlap is about 99\%) 
even though the number of degrees of freedom in the AQCM trial wave function is significantly fewer than in the AMD.

The AQCM can be applied to heavier nuclei as well. 
In $sd$-shell nuclei, a variational wave function has to include both clusters 
and quasi-clusters to describe pure shell model configurations. 
In some cases, the AQCM wave function contains more than one configuration, 
e.g. in $^{28}$Si one configuration is a pentagon of quasi-$\alpha$ clusters 
on the $xy$-plane and two $\alpha$ clusters along the $z$-axis,
and another configuration is a tetrahedron of $\alpha$ clusters 
whose center of gravity is at the origin of the coordinate system and a triangle of 
three quasi-$\alpha$ clusters on the $xy$-plane surrounding it. 
If we take the zero limit for the relative distances among $\alpha$ clusters and quasi clusters,
these two configurations become identical and give the lowest shell model configuration 
$(s_{1/2})^4(p_{3/2})^8(p_{1/2})^4(d_{5/2})^{12}$ at $\Lambda = 1$. 

A prolate shape cluster configuration $^{16}$O+$^{12}$C is also expected to play a role in the low-energy region of $^{28}$Si. In such a configuration, one can easily prepare $^{12}$C as 
the $jj$-coupling shell model wave function $(s_{1/2})^4(p_{3/2})^8$ 
using the improved AQCM parametrisation. 
The optimal configuration of $^{28}$Si can then be obtained by diagonalizing the Hamiltonian matrix comprised of these different configurations.

The AQCM studies open new perspectives in studies of cluster-shell competition 
in various particle-stable systems, including neutron-rich carbon nuclei, heavy $N\alpha$ nuclei, or hypernuclei. 
The generic explanation of nuclear clustering, its universal occurrence and properties, involves understanding of the role played by the coupling of shell model states via the decay channel(s). This coupling leads to the formation of the collective near-threshold eigenstate of the open quantum system. The coupling results in the anti-Hermitian component, and interplay between Hermitian and anti-Hermitian terms is a source of collective near-threshold phenomena in the ensemble of shell model states \cite{kyoto,forsch}. The coupling to decay channels in AQCM, which could be included through the complex scaling method, is a challenging subject to be addressed in the near future.

\begin{acknowledgments}
Two of the authors (N.I. and J.Cs.) would like to thank the bilateral program No.~119 between Japan Society for the Promotion of Science (JSPS) and Hungarian Academy of Science.
This work was supported by a Grant-in-Aid for Scientific Research No.~24$\cdot$2343 from JSPS and 
Grants Nos.~K72357, K106035 from the OTKA.
\end{acknowledgments}

\appendix

\section{$R$ expansion of the improved AQCM wave function}\label{subshell-closure}

In this Appendix, we show that the improved AQCM wave function (in this case three quasi-$\alpha$ clusters)
becomes the $jj$-coupling wave function (in this case $p_{3/2}$ subshell closure configuration) at the limit of 
$R \to 0$ and $\Lambda =1$. 
We expand the AQCM wave function in $R$ and prove that the $R^{8}$-order term of the wave function is 
the $(s_{1/2})^4(p_{3/2})^8$ configuration.

Since proton and neutron parts of the wave function are identical, we consider here only the proton part.
The proton part of the total wave function of the improved AQCM with $\Lambda = 1$ is 
the antisymmetrized product of six single particle wave functions:
\begin{widetext}
\begin{align}
	\Psi_{p} &= \left( \frac{2\nu}{\pi} \right)^{\frac{9}{2}} \mathscr{A} \Bigg[ 
		\exp \left[- \nu \left\{\bm{r}_{1} - R (1,i,0) \right\}^{2} \right] 
			\left( \begin{array}{c} 1 \\ 0 \end{array} \right), 
		\exp \left[- \nu \left\{\bm{r}_{2} - R (1,-i,0) \right\}^{2} \right] 
			\left( \begin{array}{c} 0 \\ 1 \end{array} \right), \notag \\
		&\exp \left[- \nu \left\{\bm{r}_{3} - R \left(-\frac{1}{2},i,\frac{\sqrt{3}}{2}\right) \right\}^{2} \right] 
			\left( \begin{array}{c} 1/2 \\ \sqrt{3}/2 \end{array} \right), 
		\exp \left[- \nu \left\{\bm{r}_{4} - R \left(-\frac{1}{2},-i,\frac{\sqrt{3}}{2}\right) \right\}^{2} \right] 
			\left( \begin{array}{c} -\sqrt{3}/2 \\ 1/2 \end{array} \right), \notag \\
		&\exp \left[- \nu \left\{\bm{r}_{5} - R \left(-\frac{1}{2},i,- \frac{\sqrt{3}}{2}\right) \right\}^{2} \right]
			\left( \begin{array}{c} -1/2 \\ \sqrt{3}/2 \end{array} \right),  
		\exp \left[- \nu \left\{\bm{r}_{6} - R \left(-\frac{1}{2},-i,- \frac{\sqrt{3}}{2}\right) \right\}^{2} \right] 
			\left( \begin{array}{c} -\sqrt{3}/2 \\ - 1/2 \end{array} \right) \Bigg],
		\label{total_12C}
\end{align}
\end{widetext}
where 
\begin{equation}
	(x,y,z) \equiv x \bm{e}_x + y \bm{e}_y + z \bm{e}_z
\end{equation}
stands for the Gaussian center parameters, and
\begin{equation}
	\left( \begin{array}{c} a \\ b \end{array} \right) \equiv a |\uparrow \ \rangle + b |\downarrow \ \rangle
\end{equation}
is the spin part of the wave function. Here, 
\begin{equation}
	\exp \left[- \nu \left\{\bm{r}_{1} - R (1,i,0) \right\}^{2} \right] 
	\left( \begin{array}{c} 1 \\ 0 \end{array} \right)
\end{equation}
is the spin-up proton, and
\begin{equation}
	\exp \left[- \nu \left\{\bm{r}_{2} - R (1,-i,0) \right\}^{2} \right] 
	\left( \begin{array}{c} 0 \\ 1 \end{array} \right)
\end{equation}
is the spin-down proton of the first quasi-$\alpha$ cluster. 
Single particle wave functions of other four protons are generated by rotating 
both the spatial and spin parts of these two orbits about the $y$-axis by $2 \pi / 3$ and $4 \pi / 3$.

Using Eqs.~\eqref{expansion_up}, \eqref{expansion_down}, and \eqref{convert_single_particle_wave_function},
we expand the proton part of the total wave function, Eq. \eqref{total_12C}, in $R$ as
\begin{widetext}
\begin{align}
	\Psi_{p} &= \left( \frac{2\nu}{\pi} \right)^{\frac{9}{2}} \mathscr{A} \Bigg[ 
			a_{1/2} |\frac{1}{2}, \frac{1}{2} \rangle + a_{3/2} R |\frac{3}{2}, \frac{3}{2} \rangle + O(R^{2}), 
			a_{-1/2} |\frac{1}{2}, -\frac{1}{2} \rangle + a_{-3/2} R |\frac{3}{2}, -\frac{3}{2} \rangle + O(R^{2}), \notag \\
		&a_{1/2} \left( \frac{1}{2} |\frac{1}{2}, \frac{1}{2} \rangle +\frac{\sqrt{3}}{2} |\frac{1}{2}, -\frac{1}{2} \rangle \right)
			+ a_{3/2} R \left( \frac{1}{8} |\frac{3}{2}, \frac{3}{2} \rangle + \frac{3}{8} |\frac{3}{2}, \frac{1}{2} \rangle
			+ \frac{3\sqrt{3}}{8} |\frac{3}{2}, -\frac{1}{2} \rangle 
			+ \frac{3\sqrt{3}}{8} |\frac{3}{2}, -\frac{3}{2} \rangle \right) + O(R^{2}), \notag \\
		&a_{-1/2} \left( -\frac{\sqrt{3}}{2} |\frac{1}{2}, \frac{1}{2} \rangle +\frac{1}{2} |\frac{1}{2}, -\frac{1}{2} \rangle \right)
			+ a_{-3/2} R \left( -\frac{3\sqrt{3}}{8} |\frac{3}{2}, \frac{3}{2} \rangle 
			+ \frac{3\sqrt{3}}{8} |\frac{3}{2}, \frac{1}{2} \rangle
			- \frac{3}{8} |\frac{3}{2}, -\frac{1}{2} \rangle 
			+ \frac{1}{8} |\frac{3}{2}, -\frac{3}{2} \rangle \right) + O(R^{2}), \notag \\
		&a_{1/2} \left( -\frac{1}{2} |\frac{1}{2}, \frac{1}{2} \rangle +\frac{\sqrt{3}}{2} |\frac{1}{2}, -\frac{1}{2} \rangle \right)
			+ a_{3/2} R \left( -\frac{1}{8} |\frac{3}{2}, \frac{3}{2} \rangle + \frac{3}{8} |\frac{3}{2}, \frac{1}{2} \rangle
			- \frac{3\sqrt{3}}{8} |\frac{3}{2}, -\frac{1}{2} \rangle 
			+ \frac{3\sqrt{3}}{8} |\frac{3}{2}, -\frac{3}{2} \rangle \right) + O(R^{2}), \notag \\
		&  a_{-1/2} \left( -\frac{\sqrt{3}}{2} |\frac{1}{2}, \frac{1}{2} \rangle - \frac{1}{2} |\frac{1}{2}, -\frac{1}{2} \rangle \right)
			+ a_{-3/2} R \left( - \frac{3\sqrt{3}}{8} |\frac{3}{2}, \frac{3}{2} \rangle 
			- \frac{3\sqrt{3}}{8} |\frac{3}{2}, \frac{1}{2} \rangle
			- \frac{3}{8} |\frac{3}{2}, -\frac{1}{2} \rangle 
			- \frac{1}{8} |\frac{3}{2}, -\frac{3}{2} \rangle \right) + O(R^{2}) \Bigg], \\
		&= \left( \frac{2\nu}{\pi} \right)^{\frac{9}{2}} \frac{999}{512} a_{1/2} a_{-1/2} a_{3/2}^{2} a_{- 3/2}^{2} R^{4} \mathscr{A} \Bigg[ 
		|\frac{1}{2}, \frac{1}{2} \rangle, |\frac{1}{2}, -\frac{1}{2} \rangle, |\frac{3}{2}, \frac{3}{2} \rangle,
		|\frac{3}{2}, \frac{1}{2} \rangle, |\frac{3}{2}, -\frac{1}{2} \rangle, |\frac{3}{2}, -\frac{3}{2} \rangle \Bigg]
		+ O(R^{5}).
\end{align}
\end{widetext}
The $(s_{1/2})^2(p_{3/2})^4$ configuration appears as the lowest order term in $R$, the $R^{4}$-order term.
Because of the antisymmetrization, terms up to the $R^{3}$-order vanish.
Since neutron and proton parts of the wave function have the same form, the total wave function has the $(s_{1/2})^4(p_{3/2})^8$ configuration at the $R^{8}$-order term.

\end{document}